\documentclass[sigconf,authorversion]{acmart}
\usepackage{booktabs} 
\graphicspath{{./}}
\usepackage{braket}
\usepackage{url}

\usepackage{enumerate}
\usepackage{algorithm, algorithmic}

\begin{document}
\title{Efficient and Scalable Calculation of Complex Band Structure using Sakurai-Sugiura Method}
\subtitle{(Received 3 April 2017; accepted 15 June 2017)}
\renewcommand{\shorttitle}{Efficient and Scalable Calculation of Complex Band Structure using SS Method}

\author{Shigeru Iwase}
\affiliation{%
  \institution{Department of Physics, University of Tsukuba}
  \city{Tsukuba} 
  \state{Ibaraki} 
  \postcode{305-8571}
}
\email{iwase@ccs.tsukuba.ac.jp}

\author{Yasunori Futamura}
\affiliation{%
  \institution{Master's/Doctoral Program in Life Science Innovation, University of Tsukuba}
  }
\affiliation{%
  \institution{Department of Computer Science, University of Tsukuba}
 \city{Tsukuba} 
  \state{Ibaraki} 
  \postcode{305-8573}
}
\email{futamura@cs.tsukuba.ac.jp}

\author{Akira Imakura}
\affiliation{%
  \institution{Department of Computer Science, University of Tsukuba}
 \city{Tsukuba} 
  \state{Ibaraki} 
  \postcode{305-8573}
}
\email{imakura@cs.tsukuba.ac.jp}

\author{Tetsuya Sakurai}
\affiliation{%
  \institution{Department of Computer Science, University of Tsukuba}
 \city{Tsukuba} 
  \state{Ibaraki} 
  \postcode{305-8573}
  }
\affiliation{%
  \institution{CREST, Japan Science and Technology Agency}
 \city{Kawaguchi} 
  \state{Saitama} 
  \postcode{332-0012}
}
\email{sakurai@cs.tsukuba.ac.jp}

\author{Tomoya Ono}
\affiliation{%
  \institution{Department of Physics, University of Tsukuba}
  \city{Tsukuba} 
  \state{Ibaraki} 
  \postcode{305-8571}
}
\affiliation{%
  \institution{Center for Computational Sciences, University of Tsukuba}
  \city{Tsukuba} 
  \state{Ibaraki} 
  \postcode{305-8577}
}
\email{ono@ccs.tsukuba.ac.jp}

\date{today}

\renewcommand{\shortauthors}{S. Iwase et al.}

\begin{abstract}
Complex band structures (CBSs) are useful to characterize the static and dynamical electronic properties of materials. Despite the intensive developments, the first-principles calculation of CBS for over several hundred atoms is still computationally demanding. We here propose an efficient and scalable computational method to calculate CBSs. The basic idea is to express the Kohn-Sham equation of the real-space grid scheme as a quadratic eigenvalue problem and compute only the solutions which are necessary to construct the CBS by Sakurai-Sugiura method. The serial performance of the proposed method shows a significant advantage in both runtime and memory usage compared to the conventional method. Furthermore, owing to the hierarchical parallelism in Sakurai-Sugiura method and the domain-decomposition technique for real-space grids, we can achieve an excellent scalability in the CBS calculation of a boron and nitrogen doped carbon nanotube consisting of more than 10,000 atoms using 2,048 nodes (139,264 cores) of Oakforest-PACS.
\end{abstract}

%
%

\ccsdesc[500]{Mathematics of computing~Nonlinear equations}
\ccsdesc[500]{Applied computing~Physics}

\keywords{Complex band structure, Sakurai-Sugiura method, Oakforest-PACS, Quadratic eigenvalue problem, Density functional theory, Carbon nanotube}

\copyrightyear{2017} 
\acmYear{2017} 
\setcopyright{acmcopyright}
\acmConference{SC17}{November 12--17, 2017}{Denver, CO, USA}\acmPrice{15.00}\acmDOI{10.1145/3126908.3126942}
\acmISBN{978-1-4503-5114-0/17/11}

\maketitle

\section{Introduction}
Band structure is one of the most fundamental ideas in condensed matter physics because it provides information about the electronic, magnetic, and optical properties of crystalline materials. Despite its numerous applications, band structure calculations are not always practical because actual materials of interest are not often perfect crystals, for instance; they have surfaces and interfaces. In such situations, the conventional band structure \cite{Ashcroft,Kittel:ISSP} does not give a comprehensive characterization of the properties of materials.

The complex band structure (CBS) generalizes the conventional band structure to the treatment of systems without translational symmetry by considering the imaginary components of wave vectors. The imaginary components of wave vectors describe exponentially decaying wave functions, and they allow us to predict not only static but also dynamic properties of materials, e.g., electron tunneling. The applications of CBS span a wide range of topics: surfaces and interfaces~\cite{HEINE19641,PhysRevB.30.4874,PhysRevB.24.4445}, electron transport~\cite{Fagas2004268,PhysRevB.70.045322}, magnetic tunnel junctions\cite{PhysRevLett.85.1088,PhysRevLett.95.216601}, topological materials~\cite{PhysRevB.90.155307,PhysRevB.93.115415,0953-8984-28-39-395501}, and many others. 

Regardless of the broad applications, CBS results have largely gone unrecognized since the calculation is computationally demanding for any but the simplest tight-binding approximation. First-principles calculations of CBS have so far been limited to be within a few hundred atoms. To predict the genuine properties of nano-scale materials, an efficient and robust method to calculate the CBS of over 10000 atoms is required.

The CBS computation corresponds to solving the inverse problem of that of the conventional band structure, which is obtained by choosing a real wave vector $k$ and diagonalizing $H(k)$, which yields the well-known dispersion relation $E(k)$. Here, $H(k)$ is a Bloch Hamiltonian matrix under any basis set or real-space grid. By contrast, the CBS can be obtained by putting the energy $E$ as the input and getting complex $k$ values as the output. 

To tackle this inverse problem, three methods have been widely used. The first and simplest approach is solving the determinant, $|H(k)-E|=0$ for $k$ by using the root-finding algorithm, which scans $k$ values until $|H(k)-E|$ becomes negligibly small. However, even with small systems, this approach is computationally intractable since the construction of the determinant per $k$ is time consuming and $k$ is a two-dimensional variable. Only when the $H(k)$ is given analytically this approach is useful.

The second approach uses the transfer matrix $T_{2m}(E)$ from one to another layer \cite{PhysRevB.23.4988,PhysRevB.25.3975,PhysRevB.65.165103,PhysRevB.74.205323}. Here, $m$ is the number of coupling neighbor unit cells. After constructing $T_{2m}(E)$ from the Hamiltonian, one can obtain the CBS by diagonalizing $T_{2m}(E)$. However, setting up $T_{2m}(E)$ requires the elements of the inverse matrix of the Hamiltonian, which describe the interactions with the neighbor unit cells. In addition, the eigenvalue problem for a dense matrix has to be solved to obtain the complex $k$ values. The computations of the matrix inversion and eigenvalue problem are time-consuming for large systems.

The third general approach is the M\"obius transformation method \cite{PhysRevB.55.5266,PhysRevB.83.085412}, which was developed for calculating the surface (semi-infinite) Green's function. Using the M\"obius transformation, the matrix inversion of the semi-infinite Hamiltonian can be cast as an eigenvalue problem that gives the complex $k$ values. Although the formulation gives the analytical relationship between the CBS and surface Green's function, it relies on a recursive technique of the Green's function which assumes that the Hamiltonian forms a block tri-diagonal matrix. Thus, its applications are limited to those systems that can be described within a localized-orbital basis.
	
In this paper, we present an efficient and scalable method for the first-principles calculation of CBS based on a real-space grid approach. The key idea is to express the effective single-particle equation of the bulk system as a quadratic eigenvalue problem (QEP) and compute only the solutions that are necessary for constructing the CBS by using the Sakurai-Sugiura method~\cite{Sakurai2003119,2009}. The advantage of directly solving a QEP is that we can utilize the fast techniques of conventional first-principles band calculations. For example, by using an iterative solver, we do not have to store the large sparse Hamiltonian matrix explicitly, but it suffices to multiply the Hamiltonian matrix with vectors, which speeds up the computations and reduces memory usage. Furthermore, owing to the inherently rich parallelism of the Sakurai-Sugiura method and the use of a domain-decomposition technique for real-space grids, we can achieve excellent scalability on modern massively parallel computers. Our main contributions are summarized below.

\begin{enumerate}
\item Saving memory: our method requires only $O(MN)$ memory, where $N$ is the size of the Hamiltonian matrix and $M$ is the size of the Hankel matrix in the Sakurai-Sugiura method, which is generally less than 1\% of $N$. This compares well with the $O(N^2)$ memory required by conventional methods. This small amount of memory usage allows us to perform very large scale calculations of the CBS. 

\item Time efficiency and scalability: we take the advantage of the sparsity of the Hamiltonian to iteratively solve linear equations, which are the bottlenecks of the Sakurai-Sugiura method. The sparsity of the Hamiltonian renders the matrix-vector operation useful and can scale favorably with system size. The sequential performance test shows that our method is faster than the best known method~\cite{PhysRevB.67.195315} developed for the real-space grid approach by about two orders of magnitude. In addition, our method scales very efficiently to a large number of processors owing to the hierarchical parallelism of the Sakura-Sugiura method.

\item Extension to other formalisms: our method is so versatile that one can straightforwardly extend it to other formalisms that are difficult to handle by conventional methods, such as the screened hybrid density functional~\cite{doi:10.1063/1.2187006,doi:10.1063/1.1564060}, without increasing the computational cost.
\end{enumerate}

Bulk aluminum (Al), carbon nanotube bundles, and boron- and nitrogen-doped carbon nanotubes (BN-doped CNTs) were selected as numerical examples, because bulk Al is actually used as electrode material and CNT is promising for next-generation nanodevices. Here we report the first-principles calculations of the CBS for up to 10240 atoms, an unprecedented number, using 2048 nodes (139264 cores) of Oakforest-PACS \cite{OFP}.

The rest of paper is organized as follows. Section 2 briefly introduces the formulation of CBS based on the real-space pseudopotential density functional theory method. Section 3 presents the Sakurai-Sugiura method for QEP and a parallelization technique for massively parallel computers. Sections 4 gives the serial and parallel performance analyses of our method. Section 5 describes the CBS results for carbon nanotube bundles. Finally we present  conclusions in section 6.

\begin{figure}[h]
\centering
\includegraphics[width=90mm]{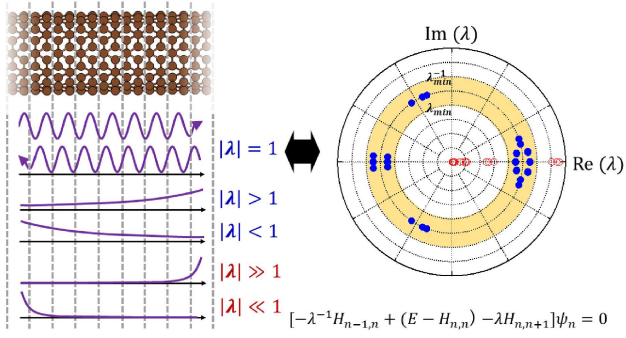}
\caption{Relationship between the CBS and solutions of QEP. Physically important eigenvalues within the green shaded area and the others are plotted as filled blue and open red dots, respectively.}
\label{fig:figure_QEP}
\end{figure}

\section{Formalism of complex band structure}
To begin, we briefly summarize the real-space pseudopotential density functional theory (DFT) method~\cite{PhysRev.136.B864,PhysRev.140.A1133,PhysRevLett.72.1240,PhysRevB.50.11355}. In the Kohn-Sham (KS) approach to DFT, the effective single-electron problem known as the KS equation is given as below

\begin{equation}
\label{eqn:KS}
H\ket{\psi}=E\ket{\psi},
\end{equation}
where $H$ is the KS Hamiltonian matrix, $E$ is the real energy, and $\ket{\psi}$ is the eigenstate. In real-space grid approach, the KS Hamiltonian matrix is highly sparse with only off-diagonal elements given by real-space finite-difference approximation for Laplacian operator and nonlocal part of pseudopotentials. For simplicity, we consider an one-dimensional crystal. Owing to the sparseness of the KS Hamiltonian, the KS equation (\ref{eqn:KS}) in $n$ th unit cell of crystalline material can be rewritten as

\begin{equation}
\label{eqn:KS2}
-H_{n,n-1}\ket{\psi_{n-1}}+(E-H_{n,n})\ket{\psi_{n}}-H_{n,n+1}\ket{\psi_{n+1}}=0,
\end{equation}
where $\ket{\psi_{n}}$ is the eigenstate of $n$ th unit cell and $H_{i,j}$ is the KS Hamiltonian matrix between the $i$ th and $j$ th unit cell. In bulk, $H_{n-1,n} = H_{n,n+1}^{\dagger}$. The Bloch's theorem yields following equations for eigenstates,
\begin{equation}
\label{eqn:bloch}
\ket{\psi_{n+l}}=\lambda^l \ket{\psi_{n}}, \quad l = -1, 0, 1,
\end{equation}
where $\lambda = e^{{\rm i}ka}$ ($a$ is the length of unit cell). By substituting \eqref{eqn:bloch} into KS equation \eqref{eqn:KS2}, one can obtain a QEP for $\lambda$ from $\lambda^{-1}$ to $\lambda$,
\begin{equation}
\label{eqn:qep}
[-\lambda^{-1}H_{n,n-1}+(E-H_{n,n})-\lambda H_{n,n+1}]\ket{\psi_{n}}=0.
\end{equation}

The dispersion of complex $k$ values obtained by scanning the energy $E$ is known as the CBS. The states $|\lambda|=1$ correspond to propagating modes. The remaining states, $|\lambda| \neq 1$, decaying or growing exponentially in real space, correspond to the evanescent modes. Evanescent modes with too small or large $|\lambda|$ are decaying or growing so rapid that they contribute marginally on the physical phenomena. Thus, it is enough to calculate the only eigenpairs of QEP \eqref{eqn:qep} satisfying

\begin{equation}
\label{eqn:criteria}
\lambda_{min}<|\lambda|<\lambda^{-1}_{min},
\end{equation}
where the order of 0.1 is enough for $\lambda_{min}$ to obtain the CBS. To find the target eigenvalues \eqref{eqn:criteria} efficiently, we employ the Sakurai-Sugiura method which will be explained in the next section.

\section{Sakurai-Sugiura method}
%
Complex moment-based eigensolvers, proposed by Sakurai and Sugiura \cite{Sakurai2003119} in 2003, compute all eigenvalues within the target region using a contour integral.
Regarding parallel computing efficiency, complex moment-based eigensolvers have a big advantage compared with Krylov-type methods because the most time-consuming part of the complex moment-based eigensolvers is the contour integral, which is suitable for parallel computing rather than the sequential procedure of the Krylov-type methods.
Based on the contour integral, the complex moment-based eigensolvers have higher level hierarchical parallelism than others.
Thanks to this high-level hierarchical parallelism, complex moment-based eigensolvers achieve higher scalability \cite{Yamazaki:2013,Kestyn:2016}.
\par
Today, there are several complex moment-based eigensolvers including direct extensions of the Sakurai-Sugiura's approach \cite{Ikegami:2008,Ikegami:2010,Ikegami:2010b,Imakura:2014,Imakura:2017} and FEAST eigensolver \cite{Polizzi:2009} and its variants \cite{Polizzi:2014,Yin:2014,Yin:2016}.
For details, we refer \cite{Imakura:2016b} and reference therein.
The Sakurai-Sugiura methods have been developed to nonlinear eigenvalue problems, unlike the FEAST eigensolver.
Therefore, in this paper, we use the Sakurai-Sugiura method using Hankel matrices \cite{2009} for solving the target QEP~\eqref{eqn:qep}.
\par
In this section, we present the basic algorithm of the Sakurai-Sugiura method.
We also introduce an improvement technique and a parallel implementation of the Sakurai-Sugiura method specialized for the target QEP \eqref{eqn:qep}.
\subsection{Basic algorithm of Sakurai-Sugiura method}
Let $N_{rh}, N_{mm} \in \mathbb{N}$ be input parameters and $V \in \mathbb{C}^{N \times N_{rh}}$ be an input matrix.
We define complex moment matrices
\begin{equation}
        S_k := \frac{1}{2 \pi {\rm i} } \oint_\Gamma z^k P(z)^{-1} V {\rm d}z, 
        \label{eq:ci}
\end{equation}
where
\begin{equation*}
        P(\lambda) = -\lambda^{-1}H_{n,n-1}+(E-H_{n,n})-\lambda H_{n,n+1}.
\end{equation*}
The Sakurai-Sugiura method and other complex moment-based eigensolvers are mathematically designed based on the properties of the complex moment matrices $S_k$.
Then, practical algorithms are derived by approximating the contour integral \eqref{eq:ci} using $N_{int}$ points of numerical integration rule:
\begin{equation}
        \widehat{S}_k := \sum_{j=1}^{N_{int}} \omega_j z_j^k P(z_j)^{-1} V
        \label{eq:numerical_integral}
\end{equation}
where $z_j$ is a quadrature point on the boundary $\Gamma$ of the target region $\Omega$ and $\omega_j$ is its corresponding weight.
The practical algorithms comprise the following three steps:
\begin{enumerate}[{Step} 1.]
\item Solve $N_{int}$ linear systems with $N_{rh}$ right-hand sides:
\begin{equation}
        P(z_j) Y_j = V, \quad
        j = 1, 2, \dots, N_{int}.
        \label{eq:blinear}
\end{equation}
\item Construct complex moment matrices, $\widehat{S}_k$ and/or others, from $Y_j$ $(j = 1, 2, \dots, N_{int})$.
\item Extract the target (approximate) eigenpairs from the complex moment matrices.
\end{enumerate}
\par
The Sakurai-Sugiura method \cite{2009} defines complex moment matrices $\widehat\mu_k = V^\dagger \widehat{S}_k$ and extracts the target (approximate) eigenpairs by solving $N_{rh} \times N_{mm} \ll N$ dimensional generalized eigenvalue problem
%
%
with block Hankel matrices
\begin{align*}
        [\widehat{T}^<]_{ij} = \widehat\mu_{i+j-1}, \quad
        [\widehat{T}]_{ij} = \widehat\mu_{i+j-2}, \quad
        1 \leq i, j \leq N_{mm}.
%
\end{align*}
\par
To reduce the computational costs and improve the numerical stability, we usually introduce a low-rank approximation with a numerical rank $\widehat{m}$ of $\widehat{T}$ based on singular value decomposition:
\begin{equation*}
        \widehat{T} 
        = [U_1, U_2] \left[
                \begin{array}{ll}
                        \Sigma_1 & O \\
                        O & \Sigma_2
                \end{array}
        \right] \left[
                \begin{array}{ll}
                        W_1^\dagger \\
                        W_2^\dagger
                \end{array}
        \right] 
        \approx U_1 \Sigma_1 W_1^\dagger.
\end{equation*}
As a result, the target QEP \eqref{eqn:qep} is reduced to an $\widehat{m}$ dimensional standard eigenvalue problem, i.e.,
\begin{equation*}
        U_1^\dagger \widehat{T}^{<} W_1 \Sigma_1^{-1} \ket{\phi} = \tau \ket{\phi}.
        \label{eq:sep_h}
\end{equation*}
The (approximate) eigenpairs are obtained as $({\lambda},\ket{{\psi_n}}) = (\tau, \widehat{S} W_1 \Sigma_1^{-1} \ket{\phi})$, where $\widehat{S} = [\widehat{S}_0, \widehat{S}_1, \dots, \widehat{S}_{N_{mm}-1}]$.
The algorithm of the Sakurai-Sugiura method is summarized in Algorithm~\ref{alg:ss-hankel}.
\begin{algorithm}[t]
\caption{Sakurai-Sugiura method for QEP}
\label{alg:ss-hankel}
\begin{algorithmic}[1]
        \REQUIRE $N_{rh}, N_{mm}, N_{int} \in \mathbb{N}, \delta \in \mathbb{R}, V \in \mathbb{C}^{N \times N_{rh}}, (z_j, \omega_j)$ for $j = 1, 2, \dots, N_{int}$
        \ENSURE $\widehat{m}$ approximate eigenpairs $({\lambda}, \ket{\psi_n})$
        \STATE Compute $\widehat{S}_k = \sum_{j=1}^{N_{int}} \omega_j z_j^k P(z_j)^{-1} V$ and $\widehat{\mu}_k = V^\dagger \widehat{S}_k$
        \STATE Set $\widehat{S} = [\widehat{S}_0, \widehat{S}_1, \dots, \widehat{S}_{N_{mm}-1}]$ and block Hankel matrices $\widehat{T}^<,\widehat{T}$\STATE Compute a low-rank approximation of $\widehat{T}$ using the threshold $\delta$: $\widehat{T}= [U_1, U_2] [\Sigma_1, O; O, \Sigma_2] [W_1, W_2]^\dagger \approx U_1 \Sigma_1 W_1^\dagger$
        \STATE Solve $U_1^\dagger \widehat{T}^{<} W_1 \Sigma_1^{-1} \ket{\phi} = \tau \ket{\phi}$, \\ and compute $({\lambda}, \ket{\psi_n}) = (\tau, \widehat{S} W_1 \Sigma_1^{-1} \ket{\phi})$
\end{algorithmic}
\end{algorithm}
\subsection{Improvement technique using the special structure of the target QEP}
\begin{figure}[t]
\begin{center}
\includegraphics[width=50mm]{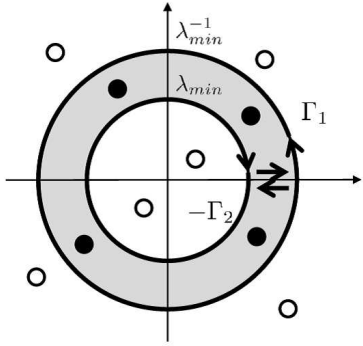}
\caption{Contour path for the target ring-shaped region. The target eigenvalues and the others are shown by $\bullet$ and $\circ$, respectively.}
\label{fig:ring}
\end{center}
\end{figure}

The target eigenvalues \eqref{eqn:criteria} are located in a ring-shaped region between two circles with center at the origin.
Radius of the small and large circles are $\lambda_{min}$ and $\lambda_{min}^{-1}$, respectively.
In this case, the contour path is set as shown in Figure~\ref{fig:ring} and the complex moment matrix $S_k$ \eqref{eq:ci} is defined by 
\begin{equation*}
        S_k := \frac{1}{2 \pi {\rm i} } \oint_{\Gamma_1} z^k P(z)^{-1} V {\rm d}z - \frac{1}{2 \pi {\rm i} } \oint_{\Gamma_2} z^k P(z)^{-1} V {\rm d}z;
\end{equation*}
see \cite{Miyata:2009}.
\par
Here, we let $N_{int}$ as the number of quadrature points for each circle.
Then, the complex moment $\widehat{S}_k$ \eqref{eq:numerical_integral} is computed by
\begin{align*}
        \widehat{S}_k = \sum_{j=1}^{N_{int}} \omega_j (z_j^{(1)})^kP(z_j^{(1)})^{-1}V 
        - \sum_{j=1}^{N_{int}} \omega_j (z_j^{(2)})^kP(z_j^{(2)})^{-1}V,
\end{align*}
where the quadrature points are
\begin{equation*}
        z_j^{(1)} =  \lambda_{min}^{-1} \exp ({\rm i} \theta_j ), \quad
        z_j^{(2)} = \lambda_{min} \exp ({\rm i} \theta_j )
\end{equation*}
and their corresponding weights are
\begin{equation*}
        \omega_j =  \frac{1}{N_{int}} \exp ({\rm i} \theta_j ),
\end{equation*}
with $\theta_j = 2 \pi (j-1/2)/N_{int}$.
The number of linear systems we need to solve is $2 \times N_{int}$, i.e.,
\begin{align}
        P(z_j^{(1)}) Y_j^{(1)} = V, 
        \label{eq:blin1} \\
        P(z_j^{(2)}) Y_j^{(2)} = V, 
        \label{eq:blin2}
\end{align}
for $j = 1, 2, \dots, N_{int}$, because the target region consists of two circles, $\Gamma_1, \Gamma_2$.
\par
The computational costs for solving the linear systems \eqref{eq:blin1} and \eqref{eq:blin2} can be reduced to half using the special structure of the target QEP \eqref{eqn:qep}.
Because $H_{n,n-1} = H_{n,n+1}^\dagger$ and $(E-H_{n,n}) = (E-H_{n,n})^\dagger$, we have
\begin{equation*}
        P(\lambda)^\dagger = P(1/\bar{\lambda}).
\end{equation*}
Therefore, the linear systems \eqref{eq:blin2} is replaced as the dual systems of \eqref{eq:blin1}, i.e.,
\begin{equation}
        P(z_j^{(1)})^\dagger Y_j^{(2)} = V,
        \label{eq:blin22}
\end{equation}
because $z_j^{(2)} = 1/\bar{z}_{j}^{(1)}$.
\par
Some linear solvers including (sparse) direct solvers and the biconjugate gradient (BiCG) method efficiently solve the linear systems \eqref{eq:blin1} and its dual systems \eqref{eq:blin22} \cite{Saad:2003}.
Specifically, the BiCG method can solve both systems with almost the same costs for solving only \eqref{eq:blin1}.
Therefore, in this paper, we select the BiCG method for solving linear systems in the Sakurai-Sugiura method.
\subsection{Parallel implementation for the target QEP}
\begin{figure}[t]
\begin{center}
\includegraphics[width=80mm]{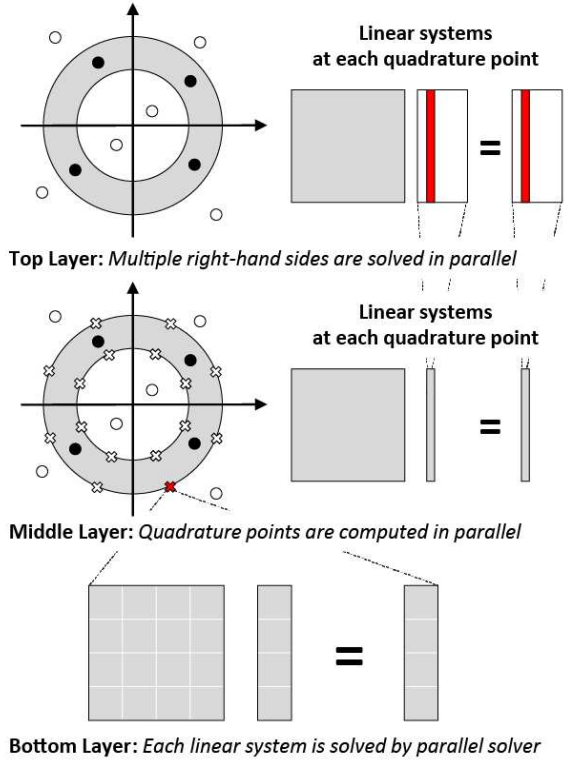}
\caption{Hierarchical parallelism of the Sakurai-Sugiura method.}
\label{fig:image}
\end{center}
\end{figure}

The most time-consuming part of the Sakurai-Sugiura method for solving QEP \eqref{eqn:qep} is Step 1 that is solving $N_{int}$ linear systems with $N_{rh}$ right-hand sides \eqref{eq:blin1} and their dual systems \eqref{eq:blin22}.
For this part, we use three layers of hierarchical parallelism of the Sakurai-Sugiura method (Figure~\ref{fig:image}).
\begin{itemize}
\item {\bf Top layer: Multiple right-hand sides}\\
As the linear systems \eqref{eq:blin1} and \eqref{eq:blin22} have $N_{rh}$ right-hand sides, we can independently solve these linear systems in $N_{rh}$ parallel without communication.
\par
This parallelism requires no communication.
Also, it is expected to have good load balancing because the convergence of the BiCG method does not strongly depend on right-hand sides.
\item {\bf Middle layer: Quadrature points}\\
As the linear systems \eqref{eq:blin1} and \eqref{eq:blin22} are independent of $j$ (index of quadrature point), we can independently solve these linear systems in $N_{int}$ parallel without communication.
\par
This parallelism requires no communication; however, we need to take care of load balancing due to the imbalance of the convergence of the BiCG method.
To achieve good load balancing, we use the following two stopping conditions for the BiCG method.
\begin{itemize}
        \item Relative residual 2-norm becomes less than certain stopping criteria.
                (This is a standard stopping condition.)
        \item The BiCG method is stopped at over half of quadrature points.
                (This is used to achieve good load balancing.)
\end{itemize}
\item {\bf Bottom layer: Each linear system}\\
The BiCG method is parallelized using domain decomposition for solving each linear system.
\par
This parallelism based on the domain decomposition requires communication in matrix-vector multiplications and inner products.
\end{itemize}
\par
The total parallelism $N_{total}$ is 
\begin{equation*}
        N_{total} = N_{dm} \times N_{int} \times N_{rh},
\end{equation*}
where $N_{dm}$ is the number of processors assigned for the domain decomposition.
If the number of processors we can use is less than $N_{int} \times N_{rh}$, we use top layer parallelism first, because upper layer is expected to show better scalability than lower layers. 
%
%

\section{Performance test}
In this section, we present a series of test calculations of the CBS. The KS Hamiltonian in the KS equation (\ref{eqn:KS}) is obtained from the electronic structure calculations using the real-space DFT code, RSPACE~\cite{PhysRevLett.82.5016,ICP}. All calculations in this section and later are performed using the norm-conserving pseudopotential proposed by Troullier and Martins~\cite{PhysRevB.43.1993}. The exchange-correlation interaction is treated by the local density approximation (LDA)~\cite{PhysRevB.23.5048} and nine-point finite-difference approximation is used for the Laplacian operator. In every case we only use the $\Gamma$ point sampling in the two-dimensional Brillouin zone and typically set a grid spacing of 0.2 angstrom.
\subsection{Serial performance}
We experimentally evaluated the serial performance of our method. Here, we considered bulk Al(100) and (6,6) armchair CNTs with 4 and 24 atoms per unit cell, respectively. The number of grid points, $N_x \times N_y \times N_z$, were $20 \times 20 \times 20$ and $72 \times 72 \times 12$, respectively. The $z$-axis was parallel to the $\langle 100 \rangle$ direction (the nanotube axis) in the case of the bulk Al(100) (CNT). We set $N_{int}=32, N_{mm}=8, N_{rh}=16, \delta = 10^{-10}$ and $\lambda_{min}=0.5 $. The convergence criteria of the BiCG method was set to $10^{-10}$. All calculations described in this subsection were carried out on a two-socket Intel Xeon E5-2683v4 with 16 cores (2.1 GHz) and 128 GB of system memory. We used the Intel Fortran compiler and Intel Math Kernel Library (MKL).

To demonstrate excellent performance of the proposed method, the computational cost was compared with that of the overbridging boundary matching (OBM) method \cite{PhysRevB.67.195315}, which is categorized as a transfer-matrix method and is the best known algorithm of the real-space grid approach. Although several improvements were proposed after the first study of the OBM method~\cite{PhysRevB.86.195406,PhysRevB.93.045421}, they did not eliminate the computations of the first and last $N_x \times N_y \times N_f$ columns of $(E-H_{n,n})^{-1}$ and the generalized eigenvalue problem for the $2 \times N_x \times N_y \times N_f$ dimensional matrices, where $N_f$ is the order of the finite-difference approximation~\cite{PhysRevB.50.11355}. In this study, the inversion matrix was calculated using the CG method, and the generalized eigenvalue problem was solved using the optimized LAPACK routine, ZGGEV.

\begin{figure}[h]
\centering
\includegraphics[width=90mm]{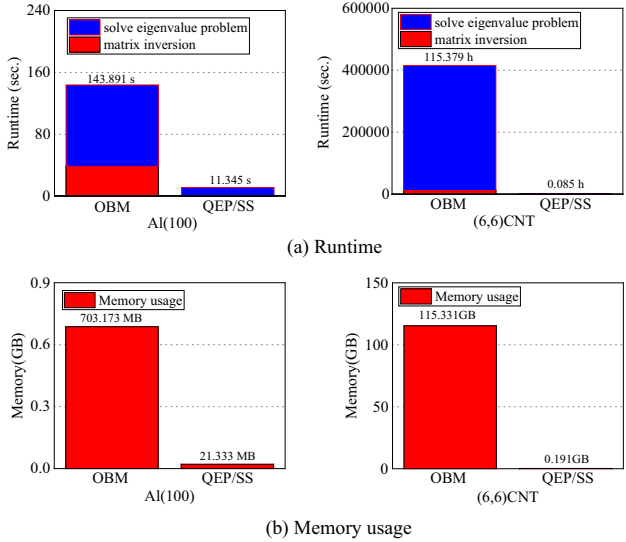}
\caption{Serial performance of the proposed (QEP/Sakurai-Sugiura) and conventional (OBM) method.}
\label{fig:serial_performance}
\end{figure}

\begin{figure}[h]
\centering
\includegraphics[width=90mm]{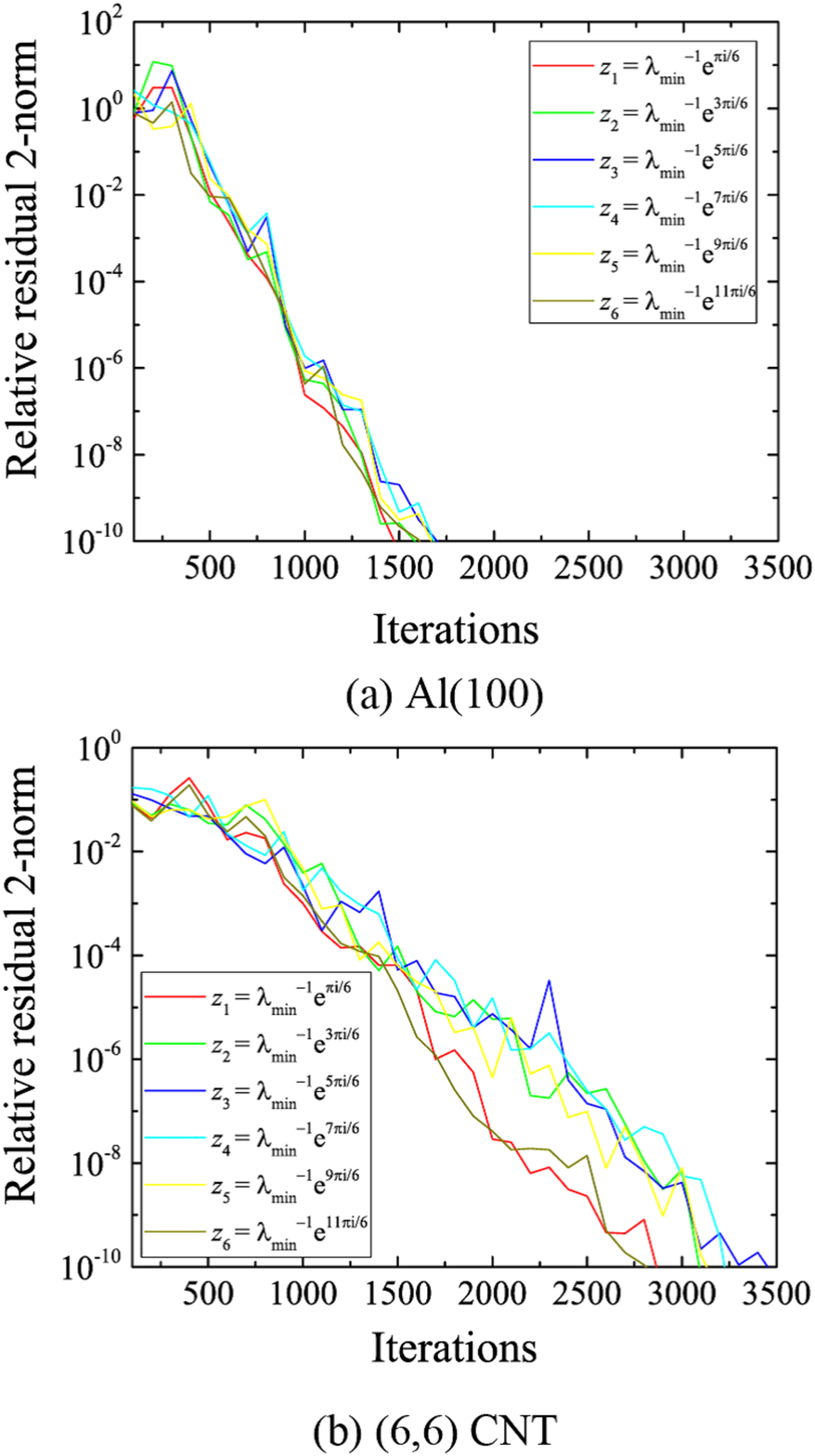}
\caption{Convergence behavior of the BiCG algorithm for (a) Al(100) and (b) (6,6) CNT at $E=E_F$. The figures show the residual norms as functions of the number of iterations at each integration point $z_j$.}
\label{fig:BiCG_conv}
\end{figure}

Figure \ref{fig:serial_performance}(a) illustrates the runtimes of the CBS calculations at $E=E_F$, where $E_F$ is the Fermi energy. In both examples, we can see that our method is surprisingly fast compared with the OBM method. We note that the solutions within $\lambda_{min}<|\lambda|<\lambda_{min}^{-1}$ obtained by our method correspond to the OBM solutions. Compared with the runtimes for Al(100), the speed-up is more prominent for (6,6) CNT. This can be seen by looking at the difference in computational complexity between the two methods. The OBM method using ZGGEV has $O(N^3)$ complexity, where $N$ is the size of the KS Hamiltonian, while our method typically has the $O(N^2)$ complexity of the Krylov subspace method until the subspace diagonalization in the Sakurai-Sugiura method becomes dominant. Figure \ref{fig:BiCG_conv} shows the histories of the residual norms of the solution vectors for linear equations at different $z_j$. In both graphs, we can see the trend that convergence does not strongly depend on the choice of $z_j$. When the half of the residual norms achieved $10^{-10}$, that with the slowest convergence  became less than $10^{-8}$. This uniform convergence behavior of the residual norms guarantees the accuracy of the present method when two stopping criteria for quadrature points as mentioned in subsection 3.3 are used. In addition, the number of iterations which need to converge generally increase as most $O(N)$ as the size of the KS Hamiltonian matrix becomes large. Actually, we can also see that the convergence of (6,6) CNT is approximately twice as slow as that of Al(100), while the matrix size for the (6,6) CNT is 7.8 times larger than that of the Al(100). Because of the sparse matrix vector operations with $O(N)$ complexity, the BiCG procedure in our method scales to less than $O(N^2)$ complexity. A huge improvement is also made on the memory requirements: a factor of 33 and 604 for Al(100) and (6,6) CNT, respectively, as shown in Figure \ref{fig:serial_performance}(b). Clearly, this improvement can be attributed to the fact that the OBM method treats dense matrices with $O(N^2)$ memory complexity, while our method treats the sparse Hamiltonian matrix with $O(N)$ memory complexity. Thus, our method is faster and more memory efficient. 

\begin{figure}[h]
\centering
\includegraphics[width=90mm]{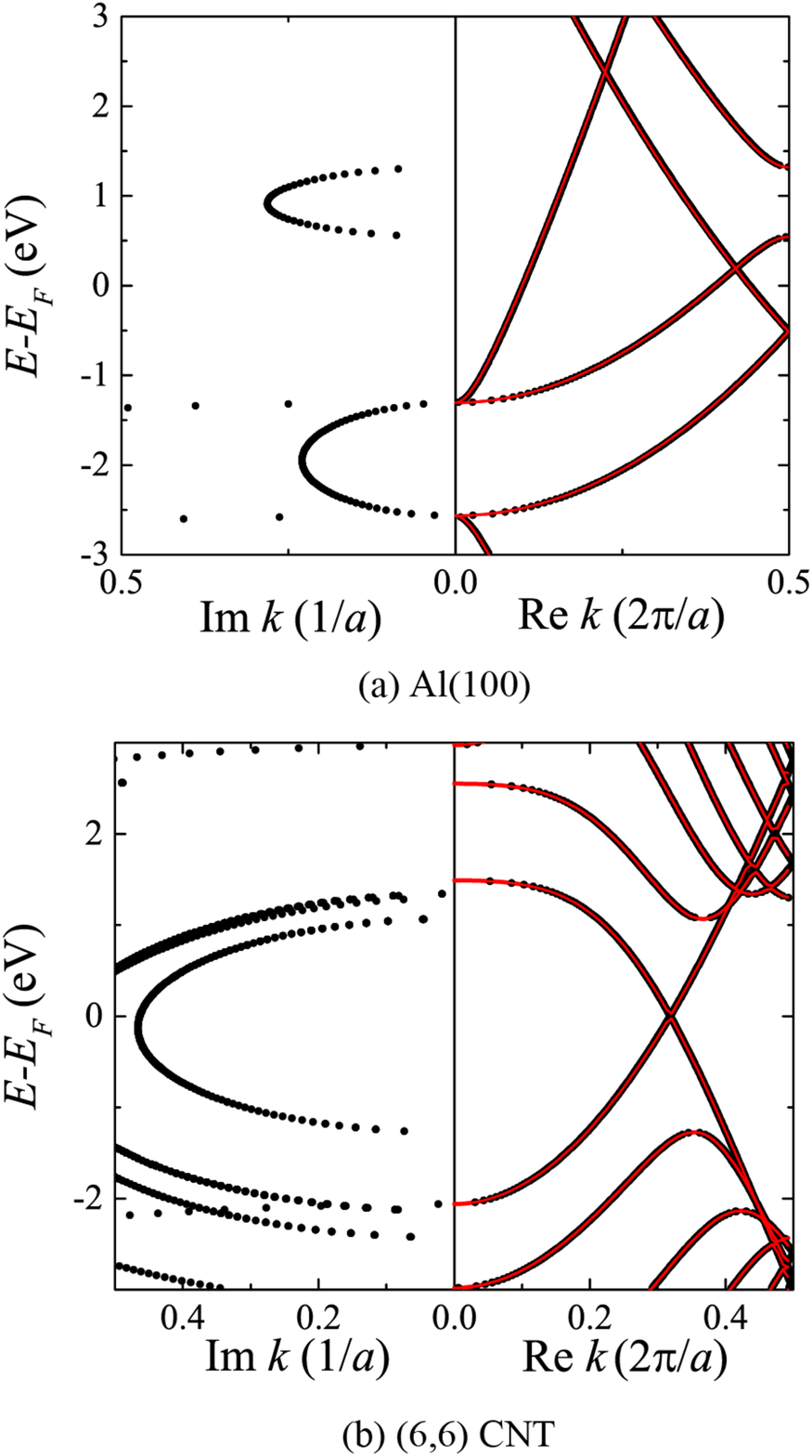}
\caption{Complex band structure for (a) Al(100) and (b) (6,6) CNT. The black dots indicate the numerical results obtained by the proposed method. The red curves show the conventionally calculated band structures for comparison.}
\label{fig:bands}
\end{figure}

Next, we evaluated the accuracy of our method. Plotting the calculated complex $k$ values which satisfy $|\lambda| = 1$ as a function of the real energy $E$ allows us to compare with the conventional band structure obtained from the electric structure calculations. As shown in Figure \ref{fig:bands}, the real $k$ values (black dots) obtained by our method are in good agreement with the conventional band structures (red curves), with an accuracy of $10^{-5}$. Moreover, our method is suitable for parallel computing, as described in the next section.

\begin{table}[h]
\centering
\caption{Breakdown of computational cost of the proposed method}
\label{table:details_calc}
\scalebox{1.2}
{
\begin{tabular}{lrr}
\hline \hline
& Al(100) & (6,6) CNT \\
\hline
read matrix data [sec.] & 0.104 & 0.209 \\
solve linear equations [sec.] & 11.207 & 304.884 \\
extract eigenpairs [sec.] & 0.138 & 0.831 \\
\hline \hline
\end{tabular}
}
\centering
\end{table}

\subsection{Parallel performance}
The computational cost of the Sakurai-Sugiura method mainly depends on the part that solves the linear equations \eqref{eq:blinear}, as shown in Table \ref{table:details_calc}. Consequently, to evaluate the parallel performance of our method, we parallelized only this part of our code by using OpenMP directives and the Intel Message Passing Interface (MPI) library. As mentioned in the previous section, the three layers of hierarchical parallelism of the Sakurai-Sugiura method, i.e., parallelisms for multiple right-hand sides (top layer), quadratic points (middle layer), and the domain decomposition (bottom layer) were introduced.

\begin{figure}[h]
\centering
\includegraphics[width=80mm]{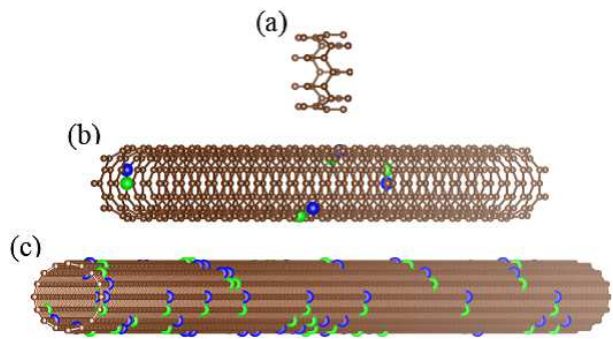}
\caption{Schematic diagrams of (a) pristine (8,0) CNT, (b) BN-doped (8,0) CNT with 1024 atoms, and (c) BN-doped (8,0) CNT with 10240 atoms. Carbon, boron, and nitrogen are depicted as brown, green, and blue balls, respectively. The BN-doped (8,0) CNTs are made by randomly inserting boron and nitrogen into a pristine (8,0) CNT.}
\label{fig:model_performance_test}
\end{figure}

All calculations in this subsection were performed on Oakforest-PACS. Each computation node is an Intel Xeon Phi$^{\rm TM}$ 7250 (code name: Knights Landing); each node has 68 cores (1.4 GHz) and 96 GB of system memory. We here conducted numerical experiments on three different (8,0) CNTs with 32, 1024, and 10240 atoms per unit cell. The computational models of the pristine (8,0) CNT with 32 atoms, the boron- and nitrogen-doped (8,0) CNT (BN-doped (8,0) CNT) with 1024 atoms, and BN-doped (8,0) CNT with 10240 atoms are shown in Figure \ref{fig:model_performance_test}. The BN-doped (8,0) CNT was made by randomly inserting boron and nitrogen into pristine (8,0) CNT.

\subsubsection{Scalability in Small System}
\begin{figure}[h]
\centering
\includegraphics[width=90mm]{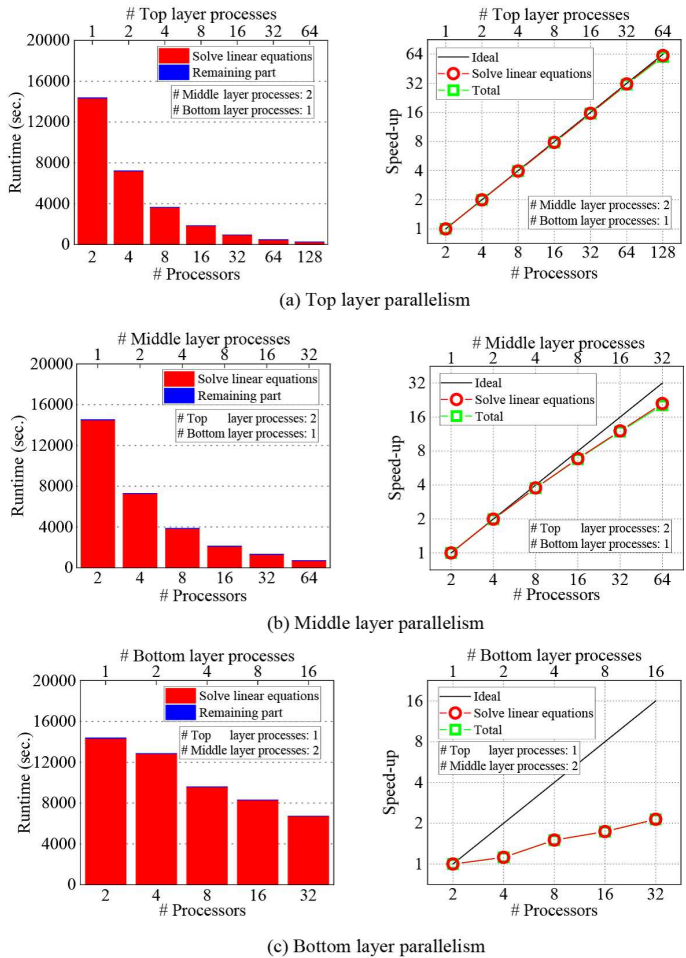}
\caption{Scalability of three layers of parallelism for (8,0) CNT with 32 atoms. 68 OpenMP threads were assigned to each MPI process.}
\label{fig:scalability_small}
\end{figure}

The system that we first tested was a (8,0) CNT with 32 atoms (the number of grid points is $72 \times 72 \times 20$). We set $N_{int}=32, N_{mm}=8, N_{rh}=64, \delta = 10^{-10}$, and $\lambda_{min}=0.5$. The convergence criteria for the BiCG method was set to $10^{-10}$. One MPI process was assigned to each node, which allowed us to use 68 OpenMP threads per MPI process. Figure \ref{fig:scalability_small} shows strong scaling of the three layers of parallelism. The total time to solve the linear equations and the time to solve the remaining part are shown on the left. Note that the remaining part does not include the disk I/O to write the solutions. The right graph in Figure \ref{fig:scalability_small} shows the speed-ups in solving the eigenvalue problem and linear equations. 

Figure \ref{fig:scalability_small}(a) shows the runtime and actual speed-up at the top layer of parallelism, where two MPI processes were assigned to the middle layer and the number of processes in the top layer was varied from 1 to 64. As we expected, the time of the remaining part is negligibly small compared with the time to solve the linear equations. Our method achieved almost ideal scaling at the top layer because the linear equations with different right-hand sides can be solved in parallel without communication and very good load-balancing can be achieved. It should be noted that the total runtime for the small system decreased from 14392 to 234 seconds when the number of processes for the top layer increased from 1 to 64.

The middle-layer scalability is shown in Figure \ref{fig:scalability_small}(b), where two MPI processes were assigned to the top layer and the number of processes in the middle layer was varied from 1 to 32. The parallel efficiency of the middle layer is slightly lower than that of the top layer, although the computations are almost independent. The degradation of scalability at the middle layer comes from the difference in convergence behaviors of the BiCG procedure at each quadrature point and it becomes more significant as the number of processes assigned to the middle layer increases. Nevertheless, the strong scaling was almost linear and a speed-up of about 21 times was achieved when we assigned 32 MPI processes to the middle layer.

Figure \ref{fig:scalability_small}(c) shows the bottom-layer scalability; here, two MPI processes were assigned to the middle layer and the number of processes in the bottom layer was varied as 1, 2, 4, 8, and 16, where the corresponding domain decompositions, $n_{x} \times n_{y} \times n_{z}$, were $1 \times 1 \times 1$, $1 \times 1 \times 2$, $1 \times 1 \times 4$, $2 \times 1 \times 4$, and $2 \times 2 \times 4$, respectively. Here $n_{x}$,  $n_{y}$, and $n_{z}$ are the number of domains in the $x$, $y$, and $z$ directions, respectively. The bottom layer scalability based on a domain-decomposition technique is much worse than the top or middle layer scalability because of frequent communications between processes in every BiCG iteration. For the small system, the poor scaling in the bottom layer is not serious because parallelization using only the top and middle layers is enough to reduce the computation time. 

We also considered how to divide the cores in the node among the OpenMP and bottom layer parallelism because it is very difficult to take full advantage of the OpenMP scalability on a many-core processor and the parallel resources are usually limited to the specific number. Table \ref{table:openmp_vs_grid} shows the elapsed times of 1000 BiCG iterations for (8,0) CNT with 32 atoms by fixing the number of cores to 64 and varying the number of OpenMP threads and $N_{dm}$. In the small system, the best performance was obtained when the 16 threads were assigned to OpenMP and four MPI processes were assigned to the bottom layer.

\begin{table*}[h]
\centering
\caption{Parallel performance inside the node. Elapsed times of 1000 iterations of the BiCG procedure for the (8,0) CNT with 32 atoms and (8,0) BN-doped (8,0) CNTs with 1024 and 10240 atoms were measured by fixing the total number of cores and splitting their allocation between the OpenMP and bottom layer parallelism.}
\label{table:openmp_vs_grid}
\scalebox{1}
{
\begin{tabular}{ccccc}
\hline \hline
 & & & Elapsed time [sec.] & \\ \cline{3-5}
\# OpenMP & \# $N_{dm}$ & (8,0) CNT (32 atoms) & BN-doped (8,0) CNT (1024 atoms) & BN-doped (8,0) CNT (10240 atoms) \\
\hline
1 & 64 & 7.77 & 104.95 & 795.42 \\
2 & 32 & 6.78 & 90.37 & 776.35 \\
4 & 16 & 5.18 & {\bf 84.77}& {\bf 774.75} \\
8 &   8 & 4.50 & 86.32 & 811.43  \\
16& 4  & {\bf 3.98} & 96.02 & 916.12 \\
32& 2 & 5.19 & 118.12 & 1132.11 \\
64& 1 & 6.16 & 161.24 & 1486.64 \\
\hline \hline
\end{tabular}
}
\centering
\end{table*}

\subsubsection{Scalability in Medium-sized System}
\begin{figure}[h]
\centering
\includegraphics[width=90mm]{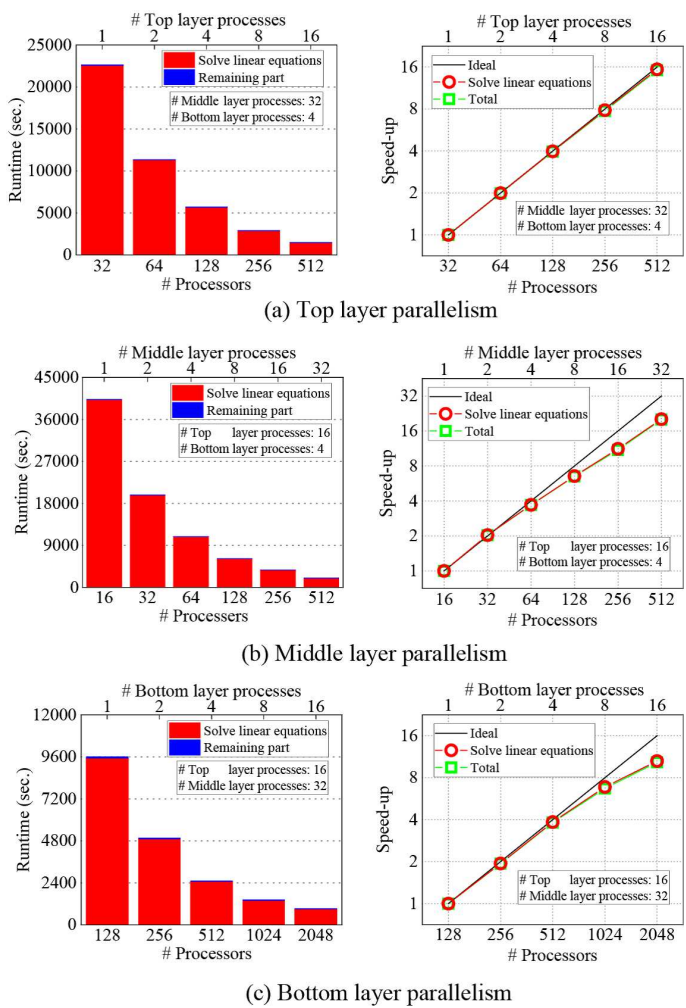}
\caption{Scalability of three layers of parallelism for BN-doped (8,0) CNT with 1024 atoms. 17 OpenMP threads were assigned to each MPI process.}
\label{fig:scalability_medium}
\end{figure}

Next, we tested the (8,0) BN-CNT with 1024 atoms (the number of grid points is $72 \times 72 \times 640$). We set $N_{int}=32, N_{mm}=8, N_{rh}=16, \delta = 10^{-10}$, and $\lambda_{min}=0.5$. The convergence criteria for the BiCG method was set to $10^{-10}$. Four MPI processes were assigned to each node, which allowed us to use 17 OpenMP threads per MPI process. Figure \ref{fig:scalability_medium}(a) shows the runtime and actual speed-up at the top layer parallelism, where 32 and 4 MPI processes were assigned to the middle and bottom layers, respectively, and the number of processes in the top layer was varied from 1 to 16. The middle-layer scalability is shown in Figure \ref{fig:scalability_medium}(b), where 16 and 4 MPI processes were assigned to the top and bottom layer and the number of processes in the middle layer was varied from 1 to 32. The bottom-layer scalability is shown in Figure \ref{fig:scalability_medium}(c), where 16 and 32 MPI processes were assigned to the top and middle layer, and the number of processes in the bottom layer was varied from 1 to 16. The domain decomposition was performed at the grid points along the $z$ direction to minimize communications and achieve better load-balancing. As shown in Figure \ref{fig:scalability_medium}, the top- and middle-layer performances are similar to the results for the small system; i.e., the top layer has the almost ideal scaling and the middle-layer scalability is slightly lower than the top-layer one. In contrast with the case of the small system, good scalability is obtained at the bottom layer. In Table \ref{table:openmp_vs_grid}, we can see that the computational time of 1000 BiCG iterations increases almost linearly relative to the number of atoms, which indicates that the communications per iteration decreases and a domain-decomposition becomes more and more efficient as the number of atoms increases. Figure \ref{fig:scalability_medium}(c) shows the CBS calculation using 2048 nodes (139264 cores) of Oakforest-PACS, i.e., 25 \% of total nodes. Even when using 2048 nodes, our method scales favorably, and the total time needed to solve the eigenvalue equation is reduced to about 905 seconds. Table \ref{table:openmp_vs_grid} shows the elapsed times of 1000 BiCG iterations for the BN-doped (8,0) CNT with 1024 atoms with 64 cores, while varying the number of OpenMP threads and $N_{dm}$. The best performance was obtained when the four threads were assigned to OpenMP and 16 MPI processes were assigned to the bottom layer.

\subsubsection{Scalability in Large System}
\begin{figure}[h]
\centering
\includegraphics[width=90mm]{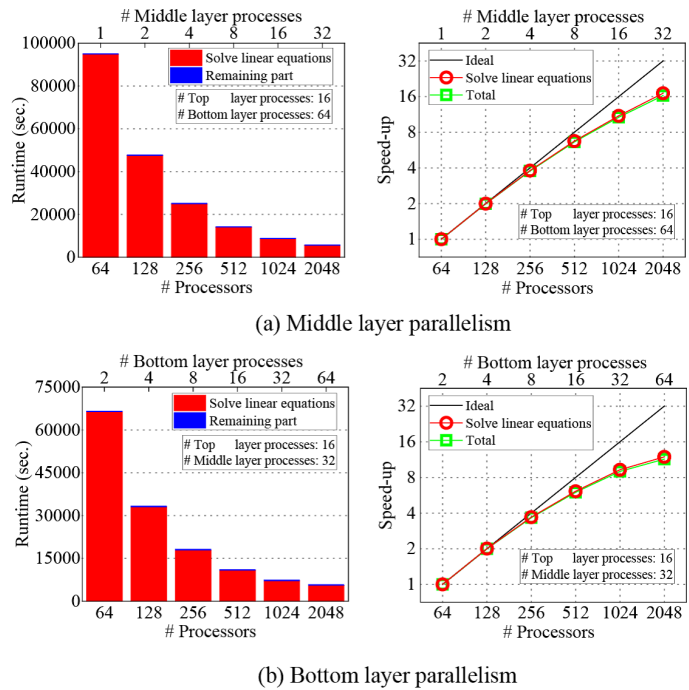}
\caption{Scalability of middle and bottom layers of parallelism for BN-doped (8,0) CNT with 10240 atoms. Four OpenMP threads were assigned to each MPI process.} 
\label{fig:scalability_large}
\end{figure}

Finally, we investigated the parallel performance on the (8,0) BN-CNT with 10240 atoms ($72 \times 72 \times 6400$ grid points). The top layer scalability is omitted due to lack of space. We set $N_{int}=32, N_{mm}=8, N_{rh}=16, \delta = 10^{-10}$, and $\lambda_{min}=0.5$. The convergence criteria for the BiCG method was set to $10^{-10}$. 16 MPI processes were assigned to each node, which allow us to use four OpenMP threads per MPI process. The domain decomposition was performed at the grid points along the $z$ direction. Figure \ref{fig:scalability_large}(a) shows the runtime of the middle layer parallelism, where 16 and 64 MPI processes were assigned to the top and bottom layer, respectively, and the number of processes for the middle layer was varied from 1 to 32. The bottom layer scalability is shown in Figure \ref{fig:scalability_large}(b), where 16 and 32 MPI processes were assigned to the top and middle layers and the number of processes in the bottom layer was varied from 2 to 64. The reduced efficiency at the large number of MPI process at bottom layer is caused by the computational cost for the global communication in the operations of nonlocal pseudopotential-vector products, which can be reduced by replacing it to local communication. Although we still need further tuning of our code, the proposed method is efficient and scalable to a large number of processors. We also note that the CBS calculations of BN-doped CNT with 10240 atoms can be executed in 2 hours using 25 \% of the computational power of Oakforest-PACS. Table \ref{table:openmp_vs_grid} shows the elapsed times of 1000 BiCG iterations for the BN-doped (8,0) CNT with 10240 atoms on 64 cores while varying the number of OpenMP threads and $N_{dm}$. The best performance was obtained when four threads were assigned to OpenMP and 16 MPI processes were assigned to the bottom layer. 

\section{Application}
\begin{figure*}[h]
\centering
\includegraphics[width=180mm]{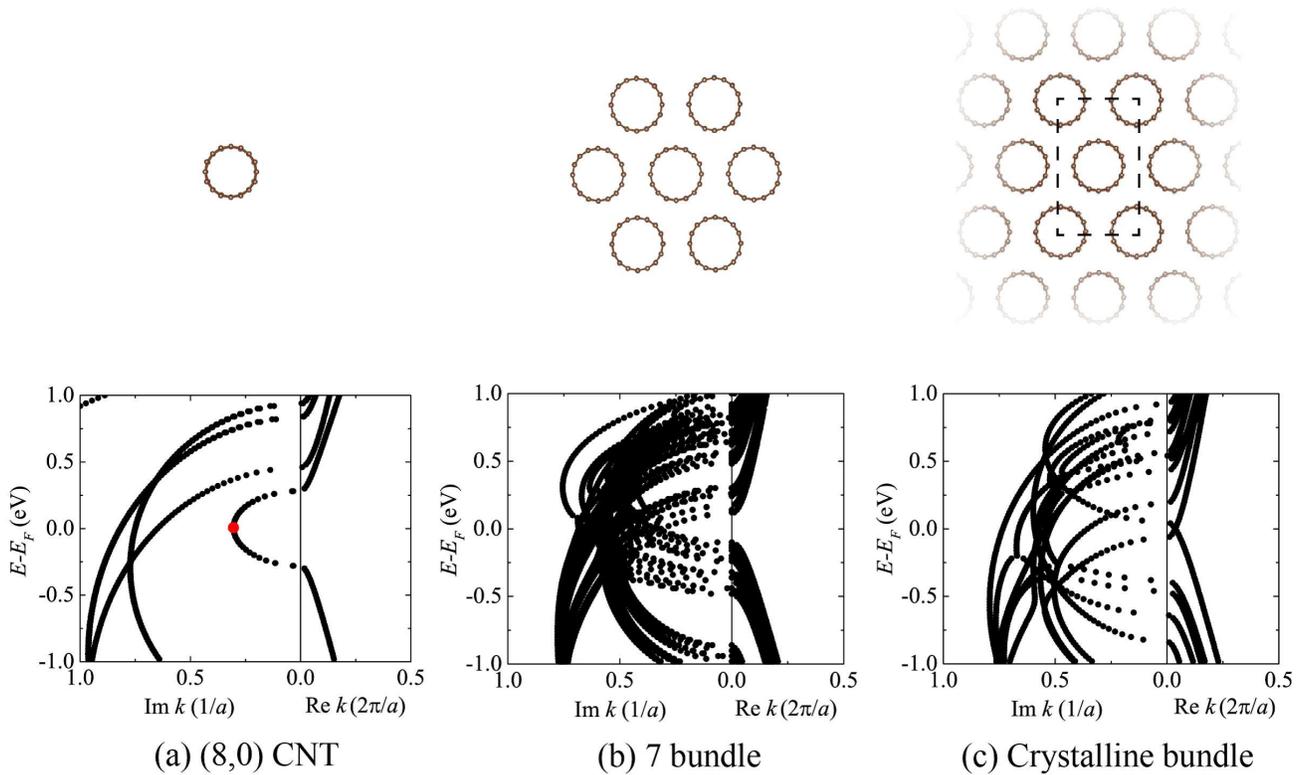}
\caption{Atomic structures and CBSs of (a) (8,0) CNT, (b) 7 bundle, and (c) crystalline bundle. Red dot in (a) denotes the branch point. Carbon atoms are brown balls. In (c), the broken line represents the boundary of unit cell for crystalline bundle.}
\label{fig:cbs_bundle}
\end{figure*}
To demonstrate the applicability of the proposed method, we perform the CBS calculation of nanotube bundles consisting of single-wall (8,0) CNTs. Although mechanical and electronic properties of bundles have been studied theoretically and experimentally so far \cite{doi:10.1143/JPSJ.70.2345,PhysRevB.84.064121,Thess483,PhysRevB.65.155411,PhysRevB.62.1643}, there is little knowledge about the CBSs of bundles. As far as we know, no attempt has been made. Up to now, calculations of the CBS for carbon nano-materials have been limited within the empirical tight-binding approximation \cite{PhysRevB.23.4988, PhysRevB.70.045322, PhysRevB.88.045403}. While the tight-binding approximation has been often used successfully for CNTs, it tends to reproduce the CBSs very poorly when a CNT comes to interact with other CNTs. Figure \ref{fig:cbs_bundle} shows the CBS of the isolated (8,0) CNT, seven, and crystalline bundles with 32, 234, and 64 atoms, respectively. Due to the strong intertube interaction, the band dispersions of bundles are enhanced and the insulator-metal transition is occurred at crystalline bundles. As for imaginary $k$ region, the loop curvatures around the Fermi energy are enlarged by bundling, and the branch point in isolated (8,0) CNT is kicked out from the band gap, which cannot be expected by conventional band structures. 

For the calculation of the CBS for 7 bundle which is the largest system in this section, it took about 1000 seconds for solving the QEP at each $E$ using the 512 nodes of Oakforest-PACS. This result shows that first-principles CBS calculations for several hundred atoms system can be performed as daily tasks using only a small amount of the total resource of Oakforest-PACS.

\section{Conclusion}
We present an efficient and scalable method for the first-principles calculation of CBS based on real-space grid approach. To take the advantage of sparsity of the KS Hamiltonian in real-space, we directly solve a quadratic eigenvalue problem for a sparse matrix by the Sakurai-Sugiura method to obtain the target eigenvalues that correspond to the CBS, instead of computing the matrix inversion and solving an eigenvalue problem for dense matrix. Experimental results on bulk Al(100) and (6,6) CNT show that the proposed method outperforms the best known method~\cite{PhysRevB.67.195315} for real-space grid approach by two orders of magnitude in both speed and memory without sacrificing accuracy. Furthermore, owing to the multiple layers of parallelism in the Sakurai-Sugiura method and a domain-decomposition technique for real-space grids, the proposed method scales very well to a large number of processors and allows us to perform unprecedented simulations to investigate the properties of material without translational symmetry. Using 2048 nodes (139264 cores) of Oakforest-PACS, we demonstrated that the simulation of BN-doped (8,0) CNTs with 10240 atoms can be executed in 2 hours. Extending the proposed method to other formalisms which are computationally intractable by conventional methods is one of the future directions. 

\begin{acks}
The authors would like to thank Y. Hirokawa and T. Boku for helpful discussion. This research was partially supported by MEXT as a social and scientific priority issue (Creation of new functional devices and high-performance materials to support next-generation industries) to be tackled by using post-K computer, and a Grant-in-Aid for JSPS
221 Research Fellow (Grant No. 16J00911). This work was supported in part by JST/CREST, JST/ACT-I (Grant No. JP-MJPR16U6), and MEXT KAKENHI (Grant No. 17K12690). The numerical calculations were carried out on the Oakforest-PACS of Joint Center for Advanced High Performance Computing.
\end{acks}


\newpage

\appendix
\section{Artifact Description}
Here, we present the artifact that accompanies our manuscript. It will support the runtime and/or memory usage results in the manuscript.

\subsection{Description}

\subsubsection{General description}

\begin{itemize}
\item Algorithms: Sakurai-Sugiura method, BiCG method, Real-space finite-difference method
\item Program: Fortran code
\item Binary: Fortran executable
\item Dataset: atomic coordinates and local potential data obtained by real-space DFT code, RSPACE (publicly not available)
\item Experiment workflow: run RSPACE; get the local potential data; solve the QEP; observe the results
\item Public available?: No
\end{itemize}

\subsubsection{Serial performance test}

\begin{itemize}
\item Compilation: ifort 18.0.1 with -O3 -r8 -qopenmp -parallel -xHost
\item Liblary link:  -Wl,--start-group \$ \{MKLROOT\} /lib/intel64/libmkl\_intel\_lp64.a \$ \{MKLROOT\}/lib/intel64/libmkl\_sequential.a \$\{MKLROOT\}/lib/intel64/libmkl\_core.a -Wl,--end-group -liomp5 -lpthread -lm -ldl
\item Hardware: Intel Xeon E5-2683v4
\end{itemize}

\subsubsection{Parallel performance test}

\begin{itemize}
\item Compilation: mpifort 17.0.2.174 with -xMIC-AVX512 -qopenmp
\item Liblary link: -mkl=parallel
\item Hardware: Intel Xeon Phi$^{\rm TM}$ 7250
\end{itemize}

\subsubsection{Application}

\begin{itemize}
\item Compilation: mpifort 17.0.2.174 with -xMIC-AVX512 -qopenmp
\item Liblary link: -mkl=parallel
\item Hardware: Intel Xeon Phi$^{\rm TM}$ 7250
\end{itemize}

\subsection{Detail of computations in Application}
In section 5 (Application), we set $N_{int}=32, N_{mm}=8, N_{rh}=64, \delta = 10^{-10}$, and $\lambda_{min}=0.5$. The convergence criteria for the BiCG method was set to $10^{-10}$. Grid points for the pristine (8,0) CNT, 7 bundle, and periodic bundle are $ 72 \times 72 \times 20$, $160 \times 148 \times 20$, and $44 \times 80 \times 20$, respectively. All calculations were executed on Oakforest-PACS. To write the complex band structures in Figure~\ref{fig:cbs_bundle}, we performed the 200 independent calculations at equidistant energies in the interval $E \in [-1{\rm eV},1{\rm eV}]$.  

\end{document}